\documentclass[final,3p,times,twocolumn]{elsarticle}

\usepackage{amssymb}
\usepackage{amsmath}
\usepackage[mathlines]{lineno}

\usepackage{lineno}

\journal{Nuclear Instruments and Methods A}

\begin{document}

\newcommand{\vekt}[1]{\mathbf{#1}}
\newcommand{\vekts}[1]{\pmb{#1}}
\newcommand{\kalman}[0]{K\'alm\'an}
\newcommand{\millepedetwo}[0]{Millepede-II}

\begin{frontmatter}



\title{General Broken Lines as advanced track fitting method}


\author{Claus Kleinwort}

\address{DESY Deutsches Elektronen-Synchrotron, Notkestra\ss e 85, 22607 Hamburg, Germany}
\ead{Claus.Kleinwort@desy.de}

\begin{abstract}
In HEP experiments the description of the trajectory of a  charged particle is obtained from a fit to measurements in tracking detectors. The parametrization of the trajectory has to account for bending in the magnetic field, 
energy loss and multiple scattering in the detector material. General broken lines implement a track model with proper description of multiple scattering leading to linear equations  with a special structure of the 
corresponding matrix allowing for a fast solution with the computing time depending linearly on the number of measurements. The calculation of the full covariance matrix along the trajectory enables 
the application to track based alignment and calibration  of large detectors with global methods. 
\end{abstract}


\end{frontmatter}


\section{Introduction}

The trajectory of a charged particle in a homogeneous magnetic field neglecting the interactions with the detector material is described by a helix.
In a global coordinate system $(x,y,z)$ with the magnetic field in $z$-direction it can be parametrized by the inverse momentum (times charge) $q/p$, an angle $\phi_0$ at and the distance $d_0$ to the point of closest approach in the $(x,y)$-plane, the dip angle $\lambda$ to that plane and the offset $z_0$ at the point of closest approach. Energy loss in the detector material 
due to ionization or radiation (for electrons) leads to a reduction of the momentum.
\emph{Multiple scattering}, mainly due to Coulomb interaction with the electrons in the atoms, results in random changes in direction and spatial position  
 with expectation values of zero and variance depending on the traversed material and the particle momentum. 
 In addition the magnetic field is usually homogeneous only in an approximation. Therefore more advanced track models are required.
 
The effect of multiple scattering can be taken into account in different ways \cite{strandlie10}. For global methods it can be added to the measurement errors leading to a non diagonal covariance matrix or explicitly fitted by scattering angles as additional track parameters. In both cases this requires the inversion of a large matrix of order $n$ (number of measurements or number of scatterers) with computing time ${\cal O}(n^3)$.
Progressive methods \cite{billoir84} like the \kalman{} filter \cite{fruehw87} updating the track parameters for each additional measurement and scatterer with a computing time ${\cal O}(n)$ have become the standard method for track fitting.

The broken lines method is a fast global track refit adding the description of multiple scattering to an initial trajectory and able 
 to determine the complete covariance matrix of all track parameters. This allows the usage as track model for track based alignment and calibration with the global \millepedetwo{} \cite{blobel06mp} algorithm.
Corrections and covariance matrices for the local track parameters (at single points) can be determined with a computing time ${\cal O}(n)$ exploiting the sparsity of the matrix of the corresponding linear equations system.
 
The original formulation \cite{blobel06bl} describes the case of a tracking detector with a solenoidal magnetic field and independent two-dimensional tracking in the bending and perpendicular to the bending plane. It constructs the 
planar trajectories from the measurements including the material around those as thick scatterers.  
In the presence of  measurements with components in both planes  a single trajectory in three dimensions  is required.
General broken lines are describing trajectories in space in an arbitrary magnetic field for arbitrary measurements and material distribution. 
The  three-dimensional broken lines trajectories from  one- or two- dimensional independent measurements dedicated to the track-based alignment of the CMS silicon tracker \cite{cmsalign11}  with \millepedetwo{} \cite{blobel06mp} are an intermediate step. They are constructed from the material between the measurement planes by equivalent thin scatterers.
Measurements are parametrized by the interpolation of (small) offsets perpendicular to the track direction defined at those scatterers and a common correction $\Delta q/p$ of the inverse momentum. Multiple scattering kinks are defined by triplets of thin scatterers.  
The corresponding linear equation system $\vekt{A} \vekt{x}=\vekt{b}$ with a bordered band matrix $\vekt{A}$ with a size $n$ depending on the number of scatterers $n_{scat}$, band width $m$ (usually 5) and border size $b=1$ is built from the Jacobians of the propagation of the offsets between measurements and scatterers on the initial trajectory and solved by root-free Cholesky decomposition.
In the following the further features of general broken lines are presented, the procedure is discussed and a comparison with the \kalman{} filter is performed. 

\section{Towards general broken lines}

\subsection{Local coordinate system}
At each measurement plane and each thin scatterer a local (orthonormal) coordinate system $(u_1,u_2,w)$
is defined. The natural choice for the $w$-direction is perpendicular to the measurement plane for a measurement  and parallel to the track direction for a scatterer.
A  local system moving with the track is the \emph{curvilinear} frame $(x_\perp,y_\perp,z_\perp)$ with $z_\perp$ in track direction and $x_\perp$ in the global $(x,y)$-plane.
At each thin scatterer the two-dimensional offset  $\vekt{u}=(u_1,u_2)$  is defined as fit parameter.

The variance $\vekt{V}_k$ of the multiple scattering kinks $\vekt{k}$ in the local system can be calculated 
from the variance $\theta_0^2\, \bigl( \begin{smallmatrix} 1 & 0 \\  0 & 1  \end{smallmatrix} \bigl)$  \cite{pdg} in the curvilinear system
by parameter transformation \cite{strandlie06} from the curvilinear to the local slopes which
 depends on the projections $c_i=\vekt{e}_\text{track}\cdot\vekt{e}_{u_i}$ of the offset directions $\vekt{e}_{u_i}$ of the local frame onto the track direction $\vekt{e}_\text{track}$:
\begin{linenomath}
\begin{multline}
 \vekt{V}_k=\frac{\partial(u'_1,u'_2)}{\partial(x'_\perp,y'_\perp)} \left( \begin{matrix} \theta_0^2 & 0 \\ 0 & \theta_0^2 \end{matrix}\right)
 \left[\frac{\partial(u'_1,u'_2)}{\partial(x'_\perp,y'_\perp)}\right]^T \negmedspace= \\
 \frac{\theta_0^2}{(1-c_1^2 -c_2^2 )^2} \left( \begin{matrix} 1-c_2^2 & c_1 c_2 \\  c_1 c_2 & 1-c_1^2  \end{matrix} \right)
\end{multline}
\end{linenomath}
With at least one offset defined perpendicular to the track direction this is a diagonal matrix. 

\subsection{Local track parameters}
In the local frame a track can be described by an offset $(u_1, u_2)$, a slope $(u'_1, u'_2)=\frac{\partial (u_1, u_2)}{\partial w}$ and the inverse momentum $q/p$. The general
broken lines fit determines from the fit parameters $\vekt{x}=(\Delta q/p,\vekt{u}_1,..,\vekt{u}_{n_{scat}})$ at each thin scatterer $i$ track parameter corrections $\Delta\vekt{p}_\text{loc}=(\Delta q/p, \vekt{u}'_i, \vekt{u}_i)$. The slope before the scattering is 
defined by $(\Delta q/p, \vekt{u}_{i-1}, \vekt{u}_i)$ and afterwards by  $(\Delta q/p, \vekt{u}_i, \vekt{u}_{i+1})$ (eqn (6) in \cite{cmsalign11}).
At the measurement planes the track parameter corrections $\Delta\vekt{p}_\text{loc}=(\Delta q/p, \vekt{u}_\text{int}', \vekt{u}_\text{int})$ can be obtained from interpolation of the enclosing scatterers (eqn (8) and inserting $\vekt{u}_\text{int}$ for $\vekt{u}_0$ in eqn (6) in \cite{cmsalign11}). In any case $\Delta\vekt{p}_\text{loc}$ depends only on $\Delta q/p$ and two adjacent offsets. Therefore for the covariance matrices of the track parameter corrections  the bordered band part of the covariance matrix $\vekt{A}^{-1}$ of the fit parameters is sufficient.
For a sparse matrix the elements of the inverse matrix inside the sparsity pattern can by calculated by special methods without those outside that pattern \cite{duff86}.
This allows to obtain the bordered band part of  $\vekt{A}^{-1}$  with ${\cal O}(n(m+b)^2)$ operations linear in the number of scatterers.

\subsection{Measurements}
The measurements $\vekt{m}$ are (the residuals with respect to the initial trajectory of) arbitrary observables with predictions defined by the local track parameters in the measurement plane.
This can be one or two dimensional  position measurements or track segments containing slope information. The linearized prediction is $\vekt{m}(\vekt{x})=\vekt{H_m} \vekt{x}$.
 In case the corresponding covariance matrices $\vekt{V}_m$ are not diagonal they should be diagonalized and the measurements and predictions modified accordingly. This allows the usage of univariate M-estimators for outlier down-weighting  and the interfacing to \millepedetwo{} \cite{blobel06mp} expecting independent scalar measurements.

\subsection{Iterations}
In case the resulting fit parameters are not small corrections as assumed for the linearization of the track model (eqn (3) in \cite{cmsalign11}) the trajectory has to be iterated. 
This requires initial non-zero values $\vekt{k}_0$ in the prediction of the multiple scattering kinks: $\vekt{k}(\vekt{x})=\vekt{H}_k \vekt{x}+\vekt{k}_0$. The expectation value remains 
$\langle \vekt{k}(\vekt{x}) \rangle=\vekt{0}$.

\subsection{Seeding}
The general broken lines are seeded by an initial trajectory. Alternatively to the result of a fit of the measurements (\emph{internal seeding}) this can be a prediction from another part of the detector (\emph{external seeding}).
The seeding track parameters at some reference point are used for the propagation along the trajectory according to the magnetic field (and average energy loss) to calculate residuals for the measurements and parameter transformation matrices. If the track fit has to be iterated to account for nonlinear effects the initial trajectory could be based on general broken lines itself.
On average no change of the track parameters at the reference point is expected from refitting: $\langle\Delta\vekt{p}_\text{seed}\rangle=\langle \vekt{H}_s \vekt{x}\rangle=\vekt{0}$.

\subsection{Minimization}
The fit parameters are determined by minimizing:
\begin{linenomath}
\begin{multline}
\chi^2(\vekt{x})=
    \sum_{i=1}^{n_\text{meas}}\left(\vekt{H}_{m,i}\vekt{x}-\vekt{m}_i\right)^T \vekt{V}^{-1}_{m,i} \left(\vekt{H}_{m,i}\vekt{x}-\vekt{m}_i\right) \\ \text{(from measurements)} \\
    +\sum_{i=2}^{n_\text{scat}-1} \left(\vekt{H}_{k,i}\vekt{x}+\vekt{k}_{0,i}\right)^T \vekt{V}^{-1}_{k,i} \left(\vekt{H}_{k,i}\vekt{x}+\vekt{k}_{0,i}\right)\:\:\text{(from kinks)}\\
    +\left(\vekt{H}_s\vekt{x}\right)^T \vekt{V^{-1}_s}\left(\vekt{H}_s\vekt{x}\right)\:\:\text{(from external seed)}
\end{multline}  
\end{linenomath}
The minimization leads to a linear equation system $\vekt{A}\vekt{x}=\vekt{b}$ with $\vekt{x}$  of size $n=2 n_\text{scat}+1$:
\begin{linenomath}
\begin{subequations}
 \begin{align}
 \vekt{A}&=\sum_{i=1}^{n_\text{meas}} \vekt{H}_{m,i}^T\vekt{V}^{-1}_{m,i}\vekt{H}_{m,i}+\sum_{i=2}^{n_\text{scat}-1} \vekt{H}_{k,i}^T\vekt{V}^{-1}_{k,i}\vekt{H}_{k,i}
 +\vekt{H}_{s}^T\vekt{V}^{-1}_s\vekt{H}_{s}\\
 \vekt{b}&=\sum_{i=1}^{n_\text{meas}} \vekt{H}_{m,i}^T\vekt{V}^{-1}_{m,i}\vekt{m}_i-\sum_{i=2}^{n_\text{scat}-1} \vekt{H}_{k,i}^T\vekt{V}^{-1}_{k,i}\vekt{k}_{0,i}
  \end{align}
\end{subequations}
\end{linenomath}

\section{Procedure}
The initial trajectory provides in the local coordinate system for the measurements the residuals $\vekt{m}$ with (diagonal) covariance matrix $\vekt{V}_m$, for the scatterers the initial (usually zero) kinks $\vekt{k}_0$ with covariance matrix $\vekt{V}_k$, optionally the external seed and 
for the parameter transformations from one point (measurement, scatterer or seed) to the next one the Jacobians $\vekt{T}_i=\partial\vekt{p}_\text{loc,i+1}/\partial\vekt{p}_\text{loc,i}$. 

The linearized track model of the general broken lines is defined at  each point by the offset part $(\vekt{d},\vekt{S},\vekt{J})_\pm=\partial\vekt{u}/\partial(q/p,\vekt{u}',\vekt{u})_\pm$ of the Jacobians for the transformtion to the next and the previous \emph{scatterer}. For the forward propagation they are calculated from the products of the point-to-point Jacobians $\vekt{T}$, for the backward case additional
partial inversions (for the offset part) are performed. In the next step the non-zero elements of the matrices $\vekt{H}$ for the linearized predictions are calculated from the $2\times2$ matrices and vectors $\vekt{J}_\pm$, 
$\vekt{W}_\pm=\pm\vekt{S}_\pm^{-1}$ and $\vekt{d}_\pm$ (see eqn (7) and (8) in \cite{cmsalign11}).
The measurements $\vekt{m}$, the kinks $\vekt{k}$, the derivatives from the sparse matrices $\vekt{H}$ and the measurement and multiple scattering errors from diagonal covariance matrices $\vekt{V}$ are the ingredients for the local fit in  \millepedetwo{} \cite{blobel06mp}.
Afterwards the linear equation system $\vekt{A} \vekt{x}=\vekt{b}$ is constructed from the individual contributions ($\vekt{H}^T\vekt{V}^{-1}\vekt{H}$, $\vekt{H}^T\vekt{V}^{-1}\vekt{m}$ or $\vekt{H}^T\vekt{V}^{-1}\vekt{k}_0$) of measurements, kinks and optionally the external seed.
The band part $\vekt{A}_u$ of the matrix is decomposed into a diagonal matrix $\vekt{D}$ and a triagonal band matrix $\vekt{L}$ with unit diagonal ($\vekt{A}_u=\vekt{L}\vekt{D}\vekt{L}^T$) and
the offsets $\vekt{x}_u$ are determined from $\vekt{D}$, $\vekt{L}$ and $\vekt{b}_u$ by solving $\vekt{L}\vekt{z}=\vekt{b}_u$ and $\vekt{L}^T\vekt{x}_u=\vekt{D}^{-1}\vekt{z}$  \cite{blobel06bl}. 
The band part of $\vekt{A}_u^{-1}$ is calculated according to \cite{duff86}. Using block matrix algebra the complete solution $\vekt{x}$ and the bordered band part of $\vekt{A}^{-1}$ are obtained.
From this corrections and covariance matrices for local track parameters at any point (measurement, scatterer or seed) can be derived.

For all this steps (propagation, prediction, construction, decomposition, solution and inversion) the number of operations is linear in the number of points or parameters and a sizable contribution to the total computing time.
As all predictions $\vekt{H}\vekt{x}$ depend at most on three consecutive  offsets the matrix $\vekt{A}_u$ has a band width $m=3\cdot2-1=5$. In case the transformation matrices $\vekt{J}$
and $\vekt{S}$ are diagonal the two offset components are decoupled and the band width reduces to $m=(3-1)\cdot2=4$ speeding up the band matrix operations (${\cal O}(n(m+b)^2)$). In addition the construction of $\vekt{A}$ is faster due to fewer non-zero elements in $\vekt{H}$.

\section{Comparision with \kalman{} filter} 

\subsection{Mathematical equivalence}
An externally seeded general broken lines trajectory with one measurement has only one offset defined and the slope has to be used 
directly as fit parameter: $\vekt{x}=\Delta\vekt{p}_{\text{loc},1}=(\Delta q/p,\vekt{u}'_1,\vekt{u}_1)$. The solution of the linear equation system is: 
\begin{linenomath}
\begin{subequations}
\begin{align}
  \vekt{x}&=\vekt{A}^{-1}\left[\vekt{H}_{m,1}^T\vekt{V}^{-1}_{m,1}\vekt{m}_1\right] \\
 \vekt{A}^{-1}&=\left[\vekt{V}^{-1}_s+\vekt{H}_{m,1}^T\vekt{V}^{-1}_{m,1}\vekt{H}_{m,1}\right]^{-1}
 \end{align}
\end{subequations}
\end{linenomath}
This is equivalent to the filtering step of  a \kalman{} filter in the weighted mean formalism (equation (8b) in \cite{fruehw87}):
\begin{linenomath}
\begin{subequations}
 \begin{align}
x_k&=\vekt{C}_k\left[\left(\vekt{C}_k^{k-1}\right)^{-1}x_k^{k-1}+\vekt{H}_k^t\vekt{V}_k^{-1}m_k \right] \\
 \vekt{C}_k&=\left[\left(\vekt{C}_k^{k-1}\right)^{-1}+\vekt{H}_k^t\vekt{V}_k^{-1}\vekt{H}_k\right]^{-1}\label{eqn-kalman}
\end{align}
\end{subequations}
\end{linenomath}
As the initial trajectory has been based on the track parameters from the external seed the prediction $x_k^{k-1}$ is zero, $\vekt{C}_k$ corresponds to
$\vekt{A}^{-1}$, $\vekt{C}_k^{k-1}$ to $\vekt{V}_s$ and $m_k$ to $\vekt{m}_1$.

In general only the covariance matrices for the track parameters at single points are calculated by the \kalman{} filter. For the global covariance matrix from all track parameters of the trajectory required 
by global alignment and calibration methods  an extension of the \kalman{} filter is described in \cite{hulsb09}.

\subsection{Computational difference}
Track fitting with the \kalman{} filter algorithm \cite{fruehw87} is a sequential procedure adding measurements and scatterers (process noise) to the trajectory one at the time.
The propagation of track parameters and covariance matrices from point to point in the filtering step involves multiplications of $5\times 5$ matrices with ${\cal O}(5^3)$ operations.
The filtering in the weighted mean formalism is dominated by the inversion of  $5\times 5$ matrices (${\cal O}(5^3)$ operations, $\vekt{C}$ in equation \eqref{eqn-kalman})   
for the addition of a single measurement. Using the gain matrix formalism requires the inversion of matrices of the size of the dimension $d$ of the measurements (${\cal O}(d^3)$ operations) 
and several multiplications of $5\times d$ matrices (${\cal O}(5^2\cdot d), {\cal O}(5\cdot d^2)$ operations). The smoothing step to get optimal local track parameters at all and not only the last point involves inversion and multiplications of $5\times 5$ matrices (${\cal O}(5^3)$ operations). For small $d$ the gain matrix formalism should be fastest. 

The general broken lines can add several ($n$, between one and all) measurements with one fit . The effort is defined by the construction of the linear equation system from the non-zero derivatives from the propagation to the previous and next thin scatterer
 and its solution  using ${\cal O}(n_{par})\cdot(m+1)^2$ operations with in general  band width $m=5$ and $n_{par}=2\,n_\text{scat}+1$, depending on the number of thin scatterers $n_\text{scat}=2..2\,n$. 

\begin{figure}
\begin{center}
\includegraphics[width=1.0\linewidth]{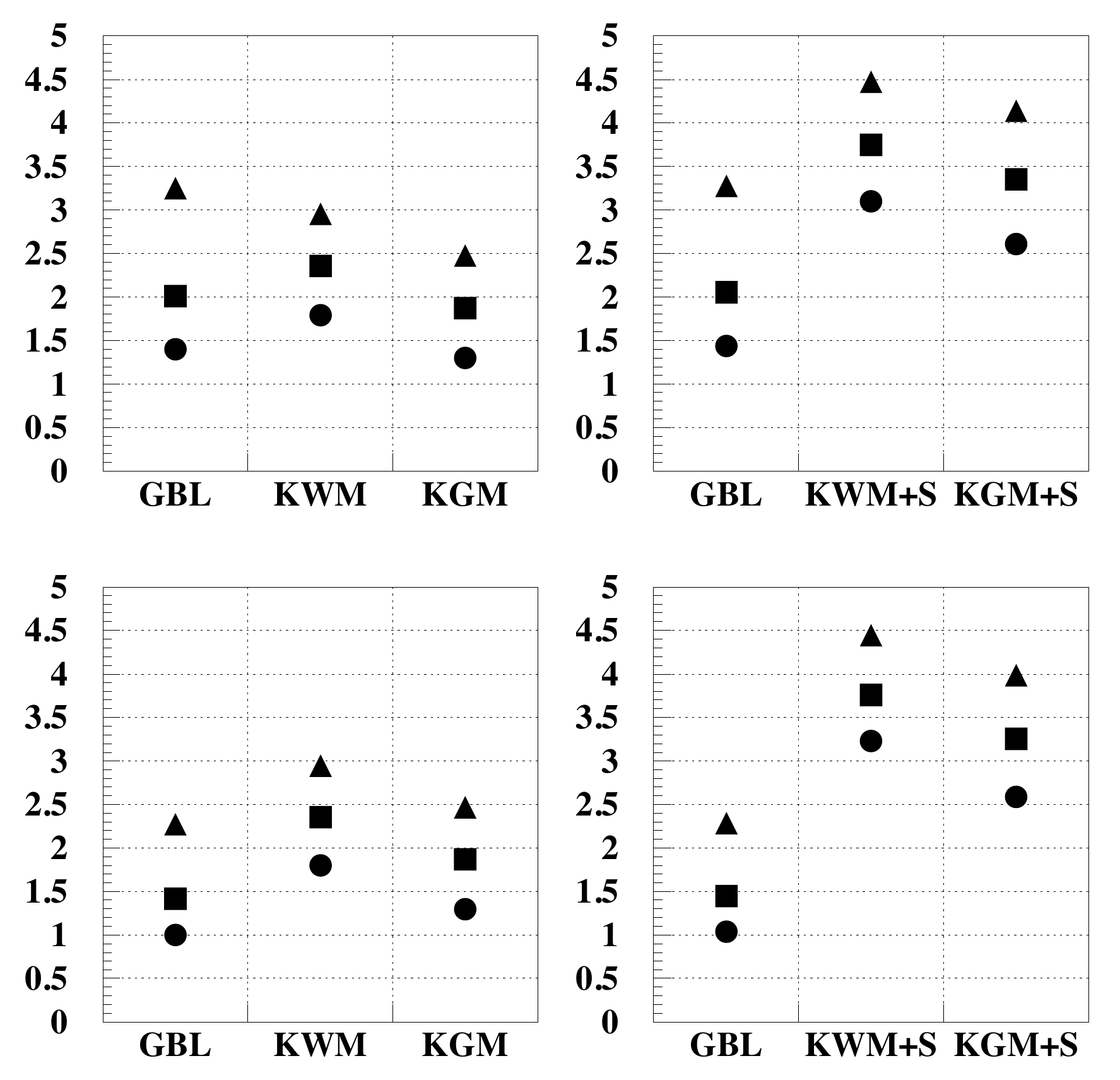} 
\end{center}
\caption{Computing time for track refitting using different algorithms, propagation Jacobians and setups of scatterers normalized to the fastest case.
The top plots are with the detailed helix propagator, the bottom ones with a simplified version, the left ones for filtering (one track end) only and the right ones
for filtering and smoothing (results at both ends). The methods shown are general broken lines (GBL) and \kalman{} filtering in weighted mean (KWM) or gain matrix (KGM) formalism. 
The case of the scattering planes coinciding with the measurement planes is indicates by circles, one thin scatterer between measurements by squares and two thin scatterers between measurements by triangles.}
\label{fig.cmpl}
\end{figure}

In a toy setup the performance of track refitting has been compared. Tracks in a detector consisting of $10$--$50$ planes with two independent measurements ($d=2$) and thin scatterers coinciding
with the measurement planes (simple model for a silicon tracker) or one or two thin scatterers between adjacent planes have been simulated. The magnetic field $\vekt{B}$ is assumed to be constant between adjacent points 
and the initial trajectory uses a detailed helix propagator \cite{strandlie06} or the simplified version for weak deflection (limit $|\vekt{B}|/p\rightarrow 0$) as in \cite{cmsalign11}.
Starting from the initial  trajectory corrections and covariance matrix of track parameters at one or both track ends have been calculated. The results of this test
are shown in figure \ref{fig.cmpl}.
For fit results at one track end \kalman{} filtering is compared with a general broken lines fit including one transformation from fit to local track parameters.
Depending on the number of scatterers per measurement the fit with general broken lines is almost as fast as the \kalman{} filter for 
the detailed propagator ($\pm 25\%$) and faster for the simplified one ($10$--$80\%$). This speedup is due to the reduced number of non-zero elements in $\vekt{H}$
for the matrix construction and reduced band width $m=4$. For the \kalman{} filter the gain matrix formalism is a little faster than the weighted mean formalism. 
For fit results at both track ends \kalman{} filtering and smoothing is compared with a general broken lines fit including two transformations from fit to local track parameters.
In this configuration the general broken lines are faster in all cases tested ($30$--$300\%$). 

\section{Summary} 
A trajectory based on General Broken Lines (GBL) is a track refit to add the description of multiple scattering to an initial trajectory based on the propagation in a magnetic field (and average energy loss).
The initial trajectory can be the result of a fit of the measurements (internal seed) or a prediction from another detector part (external seed). 
It is constructed from a sequence of (pairs of) thin scatterers describing the multiple scattering in the material between adjacent measurement planes. 
Predictions for the measurements are obtained by interpolation between the enclosing scatterers and triplets of adjacent scatterers define kink angles with variance according to the material of the central scatterer. 
The required propagation (on the initial trajectory) from a measurement plane or scatterer to the previous and next scatterer is using locally a linearized track model. 
This linearization may necessitate iterations of the fitting procedure.
The matrix of the resulting linear equations system is a bordered band matrix allowing for a fast solution by a root-free Cholesky decomposition in a time proportional to the number of measurements.
At all scatterers and measurement planes  corrections for the local track parameters are determined. The corresponding covariance matrices and the pulls for the measurements and kinks require only the calculation
of the bordered band part of the inverse matrix.

In a simulated detector, track refitting with GBL is about as fast or a little faster than \kalman{} filtering, depending on the multiple scattering setup and on the complexity of the propagation Jacobian. In comparison with \kalman{} filtering and smoothing it is up to a factor $3$ faster.

A GBL fit with external seed and one (additional) measurement is equivalent to the filtering step of the track fit with a \kalman{} filter. 

As the track refit can provide the complete covariance matrix of all track parameters GBL are well suited as track model for alignment and calibration with \millepedetwo{} \cite{blobel06mp}.





\bibliographystyle{elsarticle-num}



\end{document}